\documentclass{bioinfo}
\usepackage{array,float}
\usepackage{algpseudocode,algorithm}
\usepackage{pgffor}
\copyrightyear{2018}
\pubyear{2018}

\begin{document}
\firstpage{1}

\title[BLANT, Part 2: Extending Local Alignments]{BLANT: Basic Local Alignment of Network Topology, Part 2: Topology-only Extension Beyond Graphlet Seeds}
\author[Ding, Jain, and Hayes]{Tingyin Ding$^+$, Utsav Jain$^+$, and Wayne B. Hayes\footnote{to whom correspondence should be addressed ({\tt whayes@uci.edu}).}

\noindent $^+$ These authors contributed equally.}
\address{Department of Computer Science, University of California, Irvine CA 92697-3435, USA}

\history{Received on XXXXX; revised on XXXXX; accepted on XXXXX}

\editor{Associate Editor: XXXXXXX}

\maketitle

\begin{abstract}
BLAST is a standard tool in bioinformatics for creating local sequence alignments using a ``seed-and-extend'' approach. Here we introduce an analogous seed-and-extend algorithm that produces local {\em network} alignments: BLANT, for {\it Basic Local Alignment of Network Topology}. In Part 1, we introduced {\it BLANT-seed}, our method for generating graphlet-based seeds using only topological information. Here, in Part 2, we describe {\it BLANT-extend}, which ``grows'' seeds to larger local alignments, again using only topological information. We allow the user to specify lower bounds on several measures an alignment must satisfy, including the edge density, edge commonality (i.e., aligned edges), and node-pair similarity if such a measure is used; the latter allows the inclusion of sequence-based similarity, if desired, as well as local topology constraints. 
BLANT-extend is able to enumerate all possible alignments satisfying the bounds that can be grown from each seed, within a specified CPU time or number of generated alignments. 
While topology-driven local network alignment has a wide variety of potential applications outside bioinformatics, here we focus on the alignment of Protein-Protein Interaction (PPI) networks. We show that BLANT is capable of finding large, high-quality local alignments when the networks are known to have high topological similarity---for example recovering hundreds of orthologs between networks of the recent {\it Integrated Interaction Database} (IID). Predictably, however, it performs less well when true topological similarity is absent, as is the case in most current experimental PPI networks that are noisy and have wide disparity in edge density resulting in low common coverage.

\textbf{Availability:} {\tt https://github.com/waynebhayes/BLANT}

\textbf{Contact:} whayes@uci.edu

\textbf{Supplementary information:} None yet.
\end{abstract}

\section{Introduction}

\subsection{Motivation}

Network Alignment is the task of finding approximate isomorphisms between networks. Algorithms for network alignment can broadly be classified along three axes: (1) whether the alignments produced are local or global; (2) whether the alignments are pairwise (between exactly two networks) or multiple (between more than 2 networks simultaneously); and (3) whether the alignments are driven solely by network topology, or also include a component designed to match nodes with other types of similarity, such as sequence similarity. In this paper, we focus on the problem of {\it Pairwise Local Network Alignment} (PLNA) driven by {\em topology alone}.

Since all life on Earth is related, we expect all living cells to share some amount of similar activity in their bio-molecular interactions; in network parlance: we expect approximate isomorphisms to exist between the networks of bio-molecular interactions in the cells of of different species. Since a protein's function is essentially {\em defined} by its interaction partners, such isomorphisms can be used to facilitate cross-species prediction of protein function by relating a protein in species A with a protein in species B that is embedded in similar network topology, and is thereby likely to share similar function.

Most computational prediction of protein function today is sequence-based: some methods attempt to predict function directly from sequence (eg.,  \cite{kulmanov2020deepgoplus}), while others use longer chains of reasoning including sequence, structure, binding domains, interfaces, and compatible interaction partners (eg.,  \cite{zhang2017cofactor}). Existing functional information anywhere along this path may implicate the original protein with a similar function via the ``guilt-by-association'' principle. However, this road from sequence analysis to function prediction is long, complex, and error-prone: there are examples of proteins with no sequence similarity having near-identical function  \citep{furuse1998claudin,schlicker2006new}; examples of proteins with identical sequence having multiple, completely different functions \citep{kabsch1984use,morrone2011denatured}; and even examples of structural but not functional similarity  \citep{madsen1999psoriasis}.

In contrast, PPI interactions can be {\em directly} measured via yeast-two-hybrid and other methods \citep{VidalPPI01}, allowing direct detection of interactions and allowing a much shorter path to predictions based on guilt-by-association. Finally, since functional similarity can exist even in the absence of sequence similarity, network-based methods may be able to predict function in cases where sequence cannot. Furthermore, functions are naturally encoded using a network schema---witness how the word ``pathway'' is commonly used to describe how a set of entities coordinate their interactions to perform some function. Ergo, network information---if sufficiently complete and lacking in noise---is the more natural source of functional information. It seems natural, then, to study the patterns of network connectivity, and how these patterns can be aligned, compared, and contrasted, between species.

Unfortunately, PPI networks for most species are noisy  \citep{wodak2013protein}, incomplete \citep{vidal2016much} and biased \citep{Han2005,luck2017proteome}. 
Such data make it difficult to detect common network topology, so that ``failure to find network conservation [between] species [is] likely due to low network coverage, not evolutionary divergence.'' \citep{ideker2012differential}
For example, a recent human PPI network from BioGRID \href{https://downloads.thebiogrid.org/BioGRID/Release-Archive/BIOGRID-4.4.210}{(version 4.4.210, released June 2022)} contains 706,166 unique interactions amongst 19,662 unique human proteins; for comparison, the next most complete mammal in the same release is mouse, which contains barely 8\% of the interactions of human, at only 55,681 interactions amongst 10,221 unique mouse proteins. (Note that the numbers given on the BioGRID website for each species include {\em interactions with proteins outside the named species}. These must be removed in order to extract the PPI network of the desired species. We also remove self-interactions, to simplify the graph theory.) Given that the number of edges in the human BioGRID network has consistently grown by about 30\% each year for the past decade and shows no signs of leveling off, both networks must be considered incomplete.

Although we expect that the {\em true} PPI networks of closely-related species will be very similar, the current dearth of experimentally determined interactions makes network alignment extremely challenging. In a recent paper \citep{WeBeat} we proposed, and provided empirical evidence supporting, an Information Theoretic lower bound on the number of edges required in order for network alignments to be possible even in principle. Unfortunately, no pair of current BioGRID networks globally satisfy the lower bound---meaning that topology-only, global network alignments are impossible {\em even in principle} with current BioGRID networks.
Thus, we expect that attempting to align raw BioGRID networks will yield weak results at best.
For these reasons---and because we wish to demonstrate the efficacy of topology-only local network alignment algorithm, we will test our code on several networks that are more likely to contain truly similar topology. First, the {\it Integrated Interaction Database} (IID) \citep{kotlyar2018iid} contains partially synthetic PPI networks of 11 mammals, where the less complete networks have been augmented by edges from the more complete ones (mostly human and mouse) that are expected to be common between mammal species---for example between orthologs. Each IID network has approximately 14,000 proteins and about 300,000 edges---well above the Information Theoretic bound we proposed in \cite{WeBeat}. Since the IID networks are highly similar by construction, they form an ideal test-bed for algorithms designed to perform topology-driven network alignment. We will also test our algorithm BioGRID restricted to orthologous sub-networks, and the ``synthetic yeast'' networks of \cite{Collins2007yeast2}.

\subsection{Analogy with BLAST}

Our algorithm name---BLANT---is a clear ripoff of {\it BLAST}---the ubiquitous {\it Basic Local Alignment Search Tool} \citep{Altschul90}. The similarity in name is justified, since both BLAST and BLANT are seed-and-extend approaches. To see the analogy, we first review how BLAST works.

BLAST is an algorithm for quickly finding local alignments in genomic (or proteomic) sequences. BLAST works by first creating a comprehensive database of all $k$-letter sequences (called ``$k$-mers'') that appear in the corpus of the sequence to be searched and/or aligned. Such $k$-mers can be used to ``seed'' a local alignment between two distant regions of sequence. Below we show a hypothetical alignment between two distant regions of sequence, both of which contain the boldfaced $k$-mer (isolated for clarity):

{\tt\small\centering
\vspace{1mm}
\noindent ACTAGAT{\it C}CAC{\it C}TCTAG {\bf GAGACCGT} GTTCTTCA{\it G}AGGTG\\
\noindent ACTAGAT{\it A}CAC{\it G}TCTAG {\bf GAGACCGT} GTTCTTCA{\it T}AGGTG\\
\vspace{1mm}
}

By storing every $k$-mer and its location, BLAST can ``line up'' the regions around two identical $k$-mers, and then check to see if this local alignment ``extends'' further beyond the $k$-mers. In the case above, even though the sequences contain minor differences (highlighted with italics), the boldfaced $k$-mers form a {\it seed pair}---two identical $k$-mers, occurring in different regions, that can be used to attempt {\em extending} the match in both directions beyond the endpoints of the common $k$-mer. BLAST is extremely fast at performing the above operations, which is the reason BLAST has become the near-ubiquitous tool for comparing and aligning sequences that contain billions of letters. BLAST automatically chooses the appropriate value of $k$ to create $k$-mer seeds in a particular search and alignment task, and uses a sophisticated extend algorithm to create full seed-and-extend local alignments.

BLANT uses an analogous seed-and-extend approach, but for networks: given an undirected network $G$, and a value of $k$, BLANT samples connected, induced $k$-node subgraphs called $k$-graphlets \citep{Przulj2004Graphlets}. Since the number of $k$-graphlets in a graph of $n$ nodes is exponential in both $k$ and $n$, BLANT cannot exhaustively enumerate all $k$-graphlets, but instead samples them. If one is interested in the overall statistical properties of local topology across a network, BLANT can randomly sample millions of graphlets per second to provide a comprehensive statistical view of local network structure. While this is useful, 
our companion paper \cite{BLANT-seed} describes BLANT-seed, which takes the opposite approach: a {\em deterministic} expansion that, with care, produces $k$-graphlet seeds appropriate for expansion by the method presented in this paper.


In this paper, we describe BLANT-extend, which attempts to extend local alignments beyond the edges of a given seed consisting of two identical $k$-graphlets. While technically BLANT-extend can start with any seed---for example if $k=1$ then the two nodes could be a known pair of orthologs across two species---here we use the $k$-graphlet index produced by BLANT-seed, which uses topology alone to generate its index of $k$-graphlets. Together, BLANT-seed and BLANT-extend comprise, to the best of our knowledge, the first local alignment algorithm designed to use topology alone to generate its local alignments.

\subsection{Contribution}

So far as we are aware, there does not exist a local network alignment algorithm that uses topology alone;
most PPI network alignment algorithms today include sequence-based similarity in their objective functions; the reason is likely because, as we mentioned above, topology-only network alignments are likely {\em impossible} using current PPI network data \citep{WeBeat}.
Here, we aim to demonstrate that inter-species alignment of Protein Protein Interaction (PPI) networks is possible using topology alone, so long as enough topology is present.
We choose to omit sequence information since (a) sequence contains a known, strong signal, and including it would circumvent our goal of demonstrating the efficacy of purely topological methods that may eventually provide an independent source of protein function prediction; (b) we wish our algorithm to be applicable to any networks, not just biological ones. Additionally, once the number of experimentally determined PPIs passes the information threshold, the algorithm of this paper will likely be able to provide novel insights which cannot be found with sequence-based methods.

\section{Methods}

\subsection{Definitions}
Let $G_1,G_2$ be input graphs with nodes sets $V_1,V_2$ and edge sets $E_1,E_2$.
Let $\{u_1,u_2,u_3,\ldots\}$ be nodes from $V_1$, and let $\{v_1,v_2,v_2,\ldots\}$ be nodes from $V_2$. Assume the initial {\it seed} is a set of $k$ node pairs $S=\{(u_1,v_1),(u_2,v_2),\ldots,(u_k,v_k)\}$. We will assume that the $k$ nodes in each graph form a connected subgraph, that the induced subgraph on these $k$ nodes are identical, and that the pairs in $S$ form an initial local alignment that is ``perfect'' in the sense that an edge $(u_i,u_j)\in E_1$ if and only if the corresponding edge $(v_i,v_j)\in E_2$, for $1\le i,j \le k$. In other words, $S$ consists of two identical $k$-graphlets, perfectly aligned.

\begin{figure}
    \centering
    \includegraphics[width=0.5\textwidth]{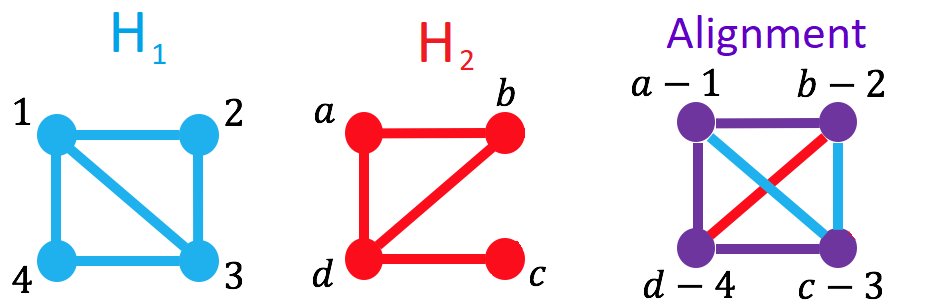}
    \caption{A schematic depiction of a 1-to-1 Pairwise Local Network Alignment. The subgraphs depicted, $H_1,H_2$ (blue and red, respectively) are induced from the original graphs $G_1,G_2$. The local network alignment can be depicted itself as a network with three types of edges: aligned edges (purple, depicting a mix of red and blue), and unaligned edges that retain their original color from either $H_1$ (blue) or $H_2$ (red). In the aligned network, the number of aligned edges is EA$=m'=3$, while E1$=m_1=5$ and E2$=m_2=4$. Thus, EC1=$|$purple edges$|$/$m_1$=3/5, EC2=$|$purple edges$|$/$m_2$=3/4, while $S^3=|$purple edges$|/|$edges of all colors in the Alignment$|$=3/6.}
    \label{fig:NetAlign}
\end{figure}

An  {\it $n$-node local alignment} $A=\{(u_1,v_1),(u_2,v_2),\ldots,(u_n,v_n)\}$ is a set of node pairs---i.e., a 1-to-1 mapping between an $n$-node subgraph $H_1$ of $G_1$ and an $n$-node subgraph $H_2$ of $G_2$. Noting that $H_1$ and $H_2$ each have exactly $n$ nodes, assume they have $m_1$ and $m_2$ edges, respectively. Finally, assume that in the specified alignment, the number of aligned edges is $m'$ (cf. Figure \ref{fig:NetAlign}). Then $EC_1(A)\equiv m'/m_1$ and $EC_2(A)\equiv m'/m_2.$ Finally, the local {\it Symmetric Substructure Score} \citep{MAGNA} is $S^3(A)=m'/(m_1+m_2-m')$. 

From a purely topological standpoint, our goal is to find local alignments that maximize some combination of $n, EC_1, EC_2$, and $S^3$---or to choose a suitable trade-off between them since increasing $n$ is likely (eventually) to decrease some or all of $EC_i$ and/or $S^3$.

Alignments are built incrementally from $S$ by expanding outwards along edges that emanate outwards from it. Given an $n$-node alignment $A$ containing node pairs drawn from subgraphs $H_1$ and $H_2$ described above, let $C_1$ be the set of nodes in $G_1$ that are adjacent to any node in $H_1$ (excluding nodes in $H_1$ itself). We say that thee nodes in $C_1$ are ``one step'' outside $H_1$; similarly, let $C_2$ be the set of nodes in $G_2$ that are ``one step'' outside $H_2$. Finally, the Cartesian product $C=C_1\times C_2$ forms the set of {\it Candidate Pairs}---ie., node pairs that could potentially be added to $A$ to increase the size of the alignment.

\subsection{Main Idea}
\begin{table*}[]
    \centering
    \begin{tabular}{|c|l|l|}
    \hline
    Requirement(s)  & Default   & Comment \\
    \hline
    ED1, ED2        & 0.1       & Lower bound of edge densities of subgraphs on $G_{1,2}$ induced by the alignment \\
    EC1, EC2        & 0.1       & Lower bounds on local EC1 and EC2 (specified separately) \\
    $S^3$           & 0.1       & Lower bound on the local {\it Symmetric Substructure Score}\citep{MAGNA} \\
    node Degree     & 1         & Exclude from $C$ any node with degree less than this (1 means no bound) \\ 
    node pair sim   & 0.1       & if provided---we use Importance similarity \citep{HubAlign} \\
    \hline
    \end{tabular}
    \caption{Required properties of a local alignment to be included in BLANT-extend's output.}
    \label{tab:requirements}
\end{table*}

Starting with the seed alignment $S$, BLANT-extend incrementally builds alignments $A$ by expanding ``outwards'' at $A$'s edges by drawing from Candidate Pairs $C$ defined above. We do so using a recursive function $\Phi(A)$ that takes the current alignment $A$ as an argument. The top-level call starts with the seed: $\Phi(S)$. At each level of recursion, $\Phi$ constructs the Candidate Pairs set $C$ based on the current alignment $A$, and then cycles through every node pair $(u,v)$ in $C$ (recall that $u\in V_1, v\in V_2$); each pair $(u,v)$ can be added to $A$ unless doing so violates any of the bounds in Table \ref{tab:requirements}, which can be user-specified. If $(u,v)$ is added to $A$, then we recursively call $\Phi(A\cup \{(u,v)\})$. We return from the current level of recursion once we have cycled through all Candidate Pairs.

The previous paragraph comprises the {\em entire} algorithm for BLANT-extend. Although it is fairly straightforward, a na\"{i}ve implementation would result in exponential run time. If run time were not a concern, then we would not mind processing the same alignment multiple times; and when cycling through candidate pairs, the order we cycle through them would be arbitrary because, if the recursion is allowed to search exhaustively, we are guaranteed to find the best possible local alignment satisfying whatever constraints that are specified (cf. Table \ref{tab:requirements}). However, it would be better to find ``good'' alignments earlier rather than later in order to provide high quality approximate answers with lower run-times. We have found several heuristics that increase the chances of finding ``good'' alignments early:
\begin{itemize}
    \item[(1)] rather than arbitrarily adding any $(u,v)$ that results in a new alignment that satisfies the requirements of Table \ref{tab:requirements}, preferentially add $(u,v)$ pairs that {\em maximize} the new value of $EC_i$ and/or $S^3$;
    \item[(2)] Related to (1), if $|C|$ is large then cycling through all Candidate Pairs to find the best pair to add can be time consuming; it may help to {\em sort} the Candidate Pairs by some heuristic that increases the chances that a ``good'' pair is found earlier rather than later.
    \item[3)] Although expensive in RAM, we store every alignment ever seen in a dictionary (implemented as a binary search tree)---even if it does {\em not} satisfy Table \ref{tab:requirements}.
\end{itemize}
The user-specified requirements for local alignments to satisfy on output are listed in Table \ref{tab:requirements}. Every alignment found to satisfy these requirements is immediately output on-the-fly. In addition, the user may specify a hard limit on number of alignments to return, or a time bound (in minutes) of CPU time; if either of these are met, the search stops and the program ends gracefully. The sorting above increases the chance that even a short run of the code is likely to find good alignments; this was verified while generating our results in that letting the code run for 24h resulted in no significant increase in the quality of alignments found within the first 60 minutes or less.

Since our SkipList of candidate node-pairs can be sorted to provide good answers more quickly, this might be a good place to include (and sort by) sequence similarity in the biological context. However, since our goal is to build alignments driven purely by topology, we have found that sorting node pairs by HubAlign's {\it Importance} similarity gives very good results.  The Importance of a node is essentially a recursively defined average degree of a node, its neighbors, its neighbor's neighbors, etc. The Importance {\it similarity} between two nodes is simply the minimum of their two Importances. We use our own implementation of Importance \citep{MamanoHayesSANA} since the original \citep{HubAlign} has biases related to deterministic tie-breaking; breaking ties randomly eliminates the bias.

In addition to optimizing the order of traversal of $C$, efficiently maintaining $C$ itself is nontrivial. First, it would be inefficient to rebuild $C$ from scratch for each call to $\Phi$, since adding $(u,v)$ to $A$ requires only the following operation to incrementally update $C$: (i) remove $(u,v)$ from $C$, and (ii) add all neighboring pairs of $(u,v)$ to $C$, except those already in $C$. Furthermore, when returning from a recursion, we must be able to easily {\em remove} $(u,v)$ from $C$ if it ceases to be ``one step'' outside $A$.
Second, if $C$ is sorted by some criterion according to point (2) above, then adding new members must insert them at the proper place in the sorted order---but a priority queue won't work because we must be able to efficiently {\em remove} $(u,v)$ even if it is not at the top of the queue.
Third, we cannot simply discard a pair $(u,v)$ from $C$ if it fails to satisfy the constraints of Table \ref{tab:requirements}, because $(u,v)$ may be ``one step'' away from multiple members of $A$, meaning that $(u,v)$ will come up again later in the search from another ``direction''---i.e., via a different recursive call to $\Phi$ using a different alignment $B$ than the current alignment $A$; thus we must {\em remember} that $(u,v)$ didn't work starting with $A$, rather than discarding it entirely. After much experimentation, we found that a {\it Skip List} was the only data structure that seemed to allow each of these operations to be performed in $O(log(n))$ time.

\begin{figure}
    \centering
    \includegraphics[width=0.4\textwidth]{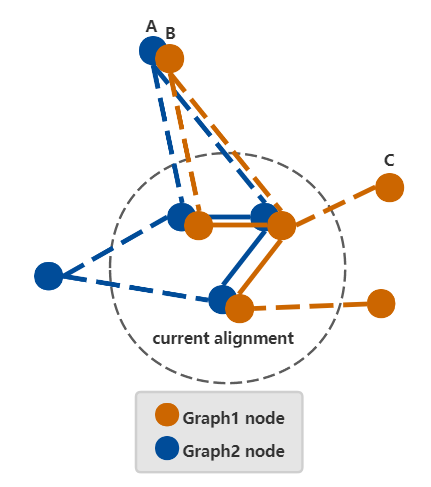}
    \caption{ There are 3 candidate nodes from graph1 and 2 candidate nodes from graph2. CandidatePairs are the combinations of the nodes from graph1 and graph2. Before aligning, E1=E2=EA=2. While aligning, M1=2 if from B to the current alignment, and M1=1 if starts from C. Similarly, M2=2 both from A. Then, if A and B are aligned, M=2 and E1=E2=EA=4. If A and C are aligned, M=1, E1=3, E2=4, and EA=3. 
    }
    \label{fig:V-being-built}
\end{figure}

Pseudo-code for the entire process is depicted in Algorithm \ref{alg:Phi}, with the following notes (cf. Figure \ref{fig:V-being-built}):
\begin{itemize}
	\item E1 and E2 are the number edges in H1 and H2 (cf. blue and red edges in Figure \ref{fig:NetAlign}), respectively.
	\item EA is the total number of aligned edge pairs (cf. purple edges Figure \ref{fig:NetAlign}) in $A$.
    \item g1node and g2node are the current nodes in consideration from G1 and G2 respectively. 
    \item M1 and M2 are the number of edges from g1node and g2node back to H1 and H2, respectively.
	\item M is the number of {\em aligned} edge pairs from (g1node,g2node) back to $H$.
	\item edgeFreq is a hashmap indexed on every $(u,v)\in C$, storing the values M1, M2, and M; these values are used by the SkipList.
\end{itemize}

\begin{algorithm}
    \caption{Alignment Algorithm}
    \label{alg:Phi}
    \begin{algorithmic}
    \Function{\sc alignStart}{Graph G1, Graph G2, Alignment seed}
        \State A = seed 
        \State allAligs = \{\} \# set of all alignments ever seen
        \State edgeFreq = \{\} \# [M,M1,M2] indexed by (g1node,g2node)
        \State SL = SkipList()  \# node pairs sorted by M in edgeFreq
        \State Cpairs = \{\}    \# CandidatePairs
        \State $\Phi$(A, Cpairs, SL)

    \EndFunction
    \\
    
    \Function{\sc $\Phi$}{Alignment A, set Cpairs, SkipList SL}
        \If{A $\not\in$ allAligs}
            \State output(A) and add it to allAligs
            \If{$|$Cpairs$|> 0$ }
                \State updateSkipList(Cpairs)
                \State seen = \{\}
                \While{$|$SL$| > 0$}
                    \State pair = SL.top() \# get but do not remove best pair
                    \If{pair $\not\in ($A $\cup$ seen) and pair satisfies Table \ref{tab:requirements}}
                        \State SL.pop() \# now {\it remove} pair from SL
                        \State A := A $\cup$ \{ pair \}
                        \State newCpairs = getNewCPairs(pair)
                        \State $\Phi$(A, Cpairs $\cup$ newCpairs, SL) 
                        \State A := A $-$ \{pair\} \# remove pair from A
                        \State Cpairs := Cpairs $\cup$ pair
                    \EndIf
                \EndWhile
            \EndIf
        \EndIf
    \EndFunction
    \\

    \Function{\sc updateSkipList}{set Cpairs}
        \For{$cPair\in Cpairs$} 
            \State M1 = numEdgesBackToH1(g1node)
            \State M2 = numEdgesBackToH2(g2node)
            \State M = numAlignedEdgesBackToH(g1node,g2node)
            \State edgeFreq[(g1node,g2node)] = [M, M1, M2]
            \State SL.add((M, (g1node, g2node)))
        \EndFor
    \EndFunction

    \end{algorithmic}
\end{algorithm}

\vspace{-3mm}\section{Results}

We applied BLANT-extend to four types of networks: \href{https://downloads.thebiogrid.org/BioGRID/Release-Archive/BIOGRID-4.4.206}{full BioGRID networks from version 4.4.206, released February 2022}; the same BioGRID but induced on known orthologs; IID version 2 \citep{kotlyar2018iid}; and the commonly used ``synthetic yeast'' dataset \citep{Collins2007yeast2}. We report our results compared to AlignMCL \citep{mina2012alignmcl}, which is the best-performing local aligner according to \cite{meng2016local}.

\subsection{Seeding}
Since seeding a local alignment is independent of extending it, we tried several types of seeds. In the case of BioGRID and IID, we can intentionally seed with known ``correct'' pairs, such as orthologs. In the case of synthetic yeast, we are aligning several different versions of the same species---{\it S. cerevisiae} (baker's yeast)---and so we can seed a node with itself. Finally, since our ultimate goal is to use topology {\em alone}---so that BLANT can be used on non-biological networks, or when no ``correct'' node mapping is known---we also use BLANT-seed \citep{BLANT-seed}.

BLANT-seed \citep{BLANT-seed} uses a sophisticated deterministic algorithm to perform $n$ independent ``walks'', starting once on each of the $n$ nodes of a given graph; on each walk, BLANT-seed will produce some number $j\ge 0$ of $k$-node graphlets ($k$-graphlets). The deterministic procedure is designed in such a way that if two networks share some regions of {\em identical} topology---that is, they contain identical, connected, induced subgraphs with significantly more than $k$ nodes each---then the same walk is likely to be performed when starting from some of the nodes in the regions of identical topology. In other words, the regions of identical topology will result in many of the same $k$-graphlets being extracted from those regions; the identical $k$-graphlets generated then act as seeds for the algorithm of this paper. BLANT-seed has been developed over several years and tested on a wide range of networks, and has been tuned to provide a reasonable trade-off between sensitivity (detecting the appropriate identical graphlets) and specificity (avoiding graphlets from disparate regions that happen, {\em by chance}, to be identical.

Although the requirement of {\em identical} topology of ``significantly'' more than $k$ nodes may seem restrictive, BLAST \citep{Altschul90} has an exactly analogous restriction with its $k$-mers. To wit, assume we are comparing two very long sequences $S_1,S_2$, and that there exist two long, highly similar regions $R_1,R_2$ with the same length, $n$---that is, they differ only by substitutions, not indels. If BLAST is using $k$-mers for some specific value of $k$, then it will detect the similarity between $R_1$ and $R_2$ only if there exists at least one common sub-sequence of length $k$ {\em in the same place} in both $R_1$ and $R_2$. (It needs to be in the same location in both $R_1$ and $R_2$, otherwise it's a false-positive seed.) So for example if BLAST is using $k$-mers with $k=20$, assume $R_1$ and $R_2$ are length 1,000 and have 95\% sequence identity. Thus, on average, one in 20 letters differ between them. Normally BLAST would have no problem discovering two sequences with 95\% similarity. However, an adversary could foil BLAST's attempt to discover the high similarity between $R_1$ and $R_2$ by methodically placing substitution errors precisely every 20 letters, with 19 ``correct'' letters between each substitution. Since BLAST is indexing $20$-mers, and the adversary has methodically ensured that there are no contiguous sequences of more than 19 correct letters, we are guaranteed that BLAST will {\em never} find two identical correctly-placed $20$-mers; and in the absence of correctly placed 20-mer seeds, BLAST has nothing useful to extend, and so it will never ``notice'' the high sequence identity between $R_1$ and $R_2$.

The primary difference between BLAST and BLANT in the above analogy is that BLAST requires only {\em one} contiguous sub-sequence of $k$ correctly-placed, identical letters to ensure a high likelihood of finding the longer region of high similarity; BLANT, on the other hand, not only requires the existence of identical $k$-graphlets in the correct place: it also requires our deterministic algorithm to {\em find} such identical $k$-graphlet(s) among the exponential number of other possibilities in the region. This is why we say that the identical regions must, ideally, have significantly more than $k$ nodes for BLANT-seed to have a good chance of finding identical, correctly-placed $k$-graphlets.

All alignments below were given one hour of CPU time. Given that BLANT-extend is currently written in Python, and that we plan to eventually convert it into C, these runtimes can be significantly improved.

\subsection{IID rat and mouse}

\begin{table}
\centering\small
\caption{Networks used in this paper}\label{tab:networks}
\begin{tabular}{|rrrrl|}
\hline
nodes   & edges & degree        & density       & name \\
\hline
15740   & 283631        & 36.04 & 0.0023        & IIDrat \\
17529   & 330304        & 37.69 & 0.0022        & IIDmouse \\
\hline
\end{tabular}
\end{table}

The IID networks are, by far, the largest PPI networks available. Although they are partly synthetic, they are currently the best available approximation to ``real'' PPI networks that (a) are nontrivial in size, and (b) share what is believed to be about the same amount of topological similarity as we expect in the (currently unknown) true networks. Statistics for the IID networks for rat ({\it R. norvegicus}) and mouse ({\it M. musculus}) are presented in Table \ref{tab:networks}.

\begin{figure}
    \centering
    \includegraphics[width=0.5 \textwidth]{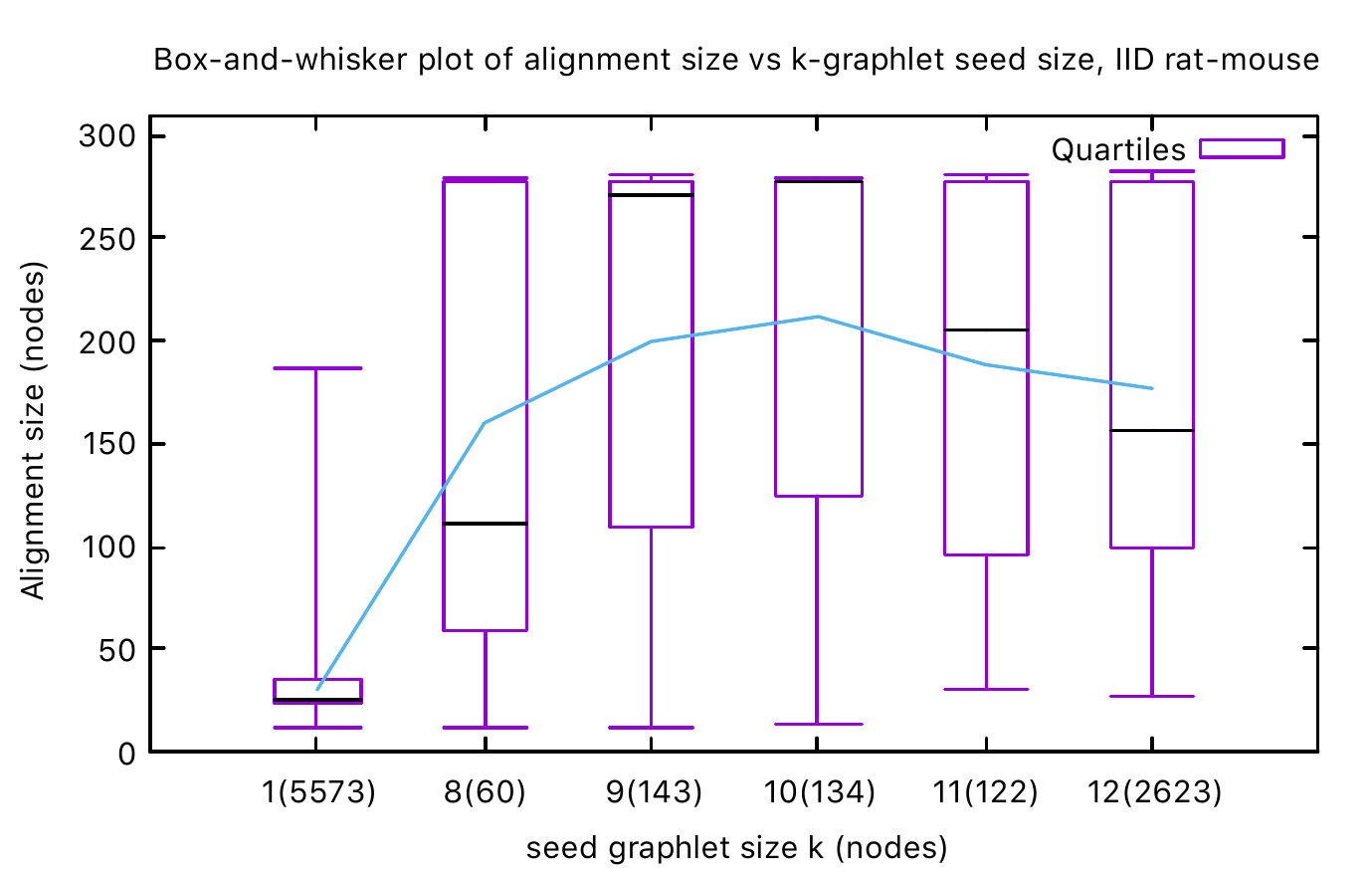}    
    \caption{{\bf Alignment size (nodes) vs. $k$-graphlet seed size for IID rat-mouse:} The horizontal axis depicts $k(n_k)$, where $k$ is the size of the seed graphlet and $n_k$ is the number of distinct seeds produced by BLANT-seed for that $k$. For each seed, we plot the maximal alignment size, in nodes, that BLANT-extend was able to achieve: the the top and bottom of each box represent the lower and upper quartile of alignment size across the $n_k$ seeds, while the line in the middle is the median. The lower and upper whiskers are the smallest and largest alignments, respectively. The cyan line connects the mean values for each $k$.}
    \label{fig:size-vs-k}
\end{figure}

Figure \ref{fig:size-vs-k} displays the maximal sized alignment, in nodes, produced by BLANT-extend across the seeds produced by BLANT-seed. The single most important observation of the plot is that $k$-graphlet seeds with $k>1$ {\em far} outperform seeds of size 1 (eg., a seed consisting of a single ortholog pair): for $k=1$, the mean, median, and even upper quartile of alignment size never exceeds 40 nodes---in fact, out of the 5573 seeds used, only two had more than 100 nodes: the maximum was 186; the second largest was 103; and all others had fewer than 70 nodes. In stark contrast, when using $k$-graphlets of size 8 or more, the {\em lower} quartile for $k\ge 8$ is higher than the {\em upper} quartile for $k=1$. (Seed graphlets with fewer than 8 nodes perform poorly, for technical reasons explained in the BLANT-seed paper \citep{BLANT-seed}.) Of particular interest is the fact that the largest alignments seem to come from $k=9$ and $k=10$, both of which have medians above 250 nodes. On the other hand, BLANT-seed was only able to produce 143 and 134 seeds, respectively, while for $k=12$ it was able to produce 2623 seeds---much higher volume, though with a median alignment size of about 150. There is clearly a trade-off here that is worth further study.

\begin{figure}
    \centering
    \includegraphics[width=0.5 \textwidth]{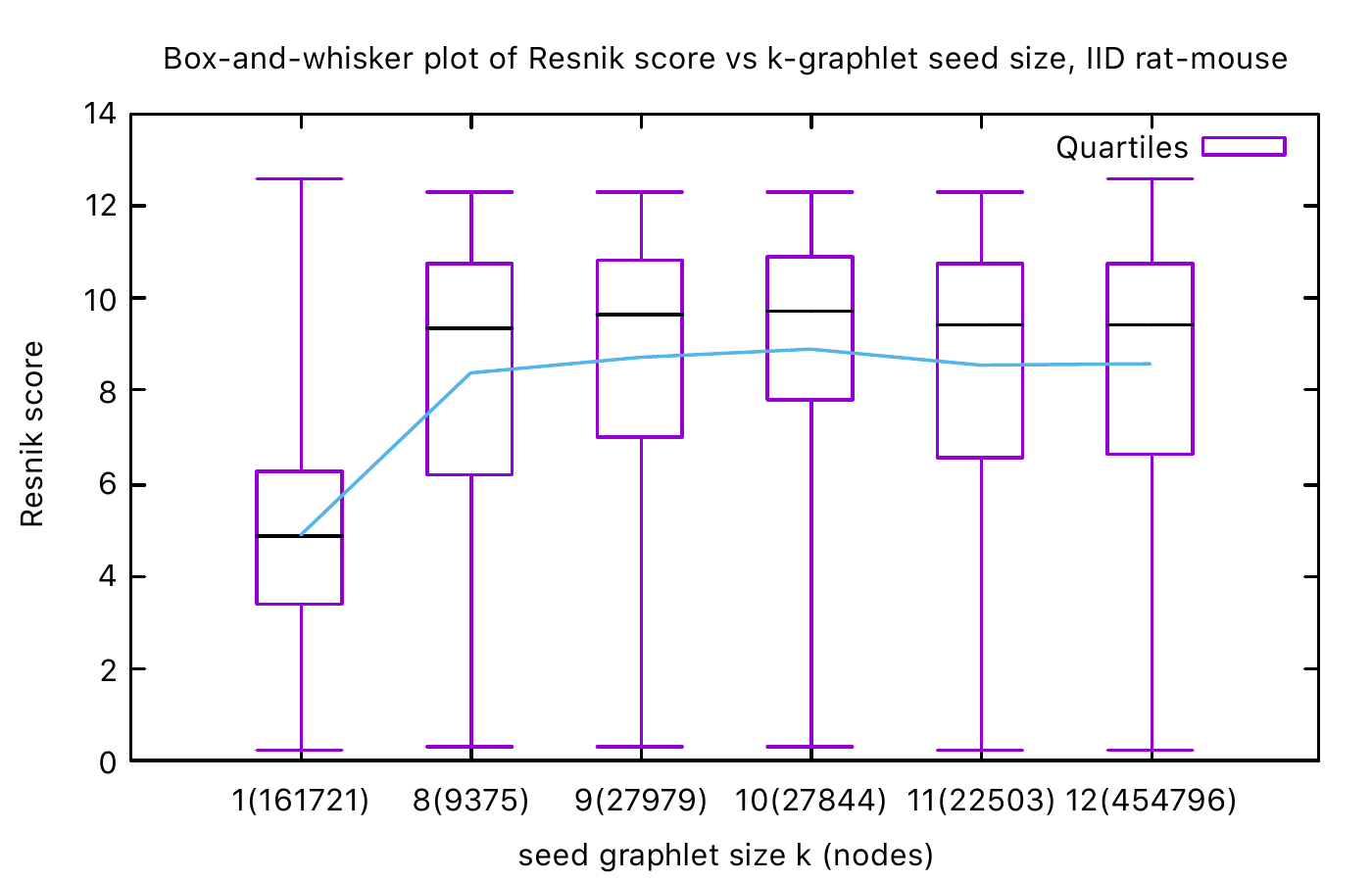}    
    \caption{{\bf Resnik semantic similarity vs. $k$-graphlet seed size for IID rat-mouse:} Similar to Figure \ref{fig:size-vs-k}, except the vertical axis is now Resnik semantic similarity (irrespective of alignment size).}
    \label{fig:resnik-vs-k}
\end{figure}

\begin{figure}
    \centering
    \includegraphics[width=0.5 \textwidth]{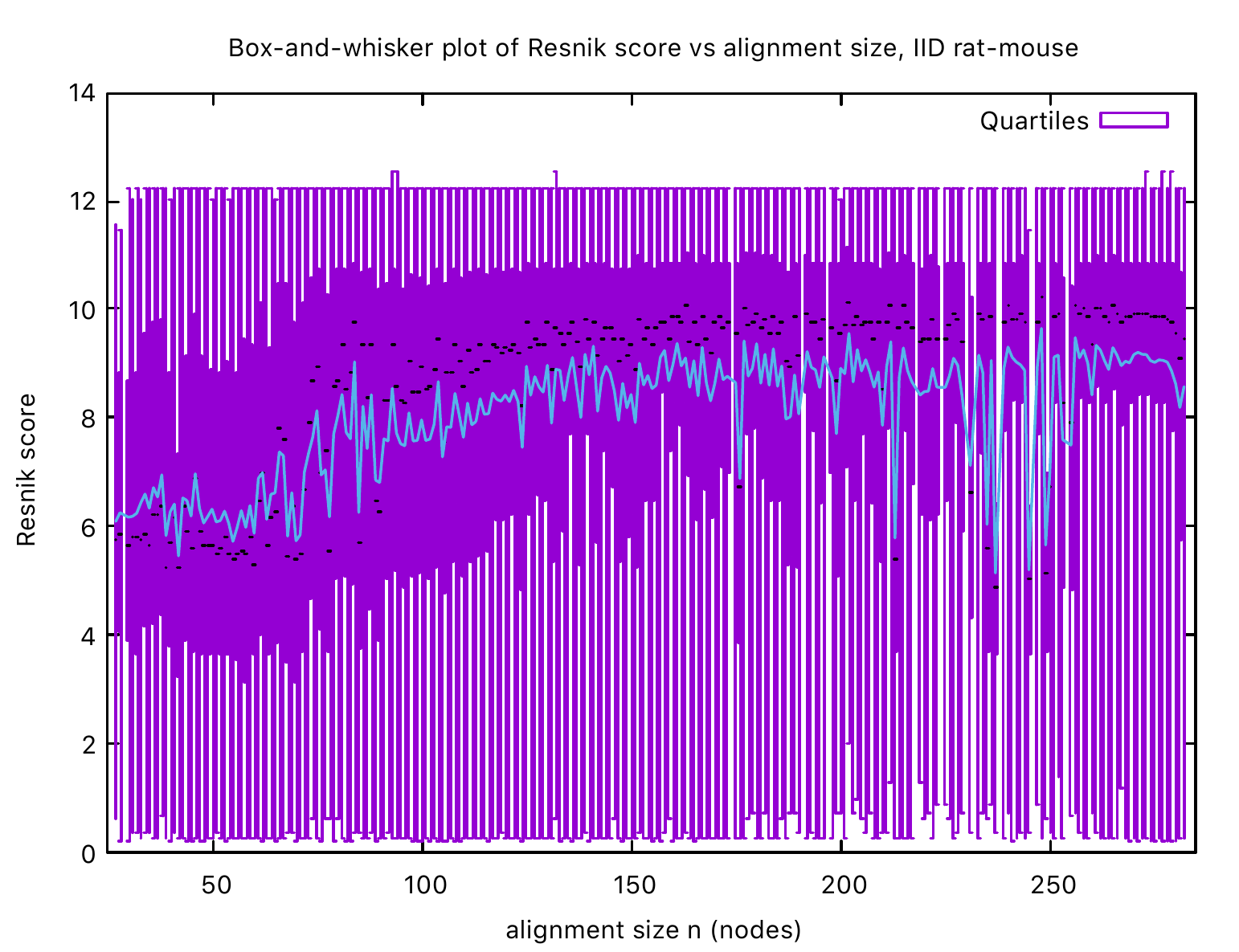}    
    \caption{{\bf Resnik semantic similarity vs. alignment size for IID rat-mouse:} Similar to Figure \ref{fig:resnik-vs-k}, except the horizontal axis is now alignment size (omitting $k=1$).}
    \label{fig:resnik-vs-size-no1}
\end{figure}

Figures \ref{fig:resnik-vs-k} and \ref{fig:resnik-vs-size-no1} display the mean Resnik scores of the alignments as a function of, respectively, $k$ irrespective of alignment size, and size irrespective of $k$---except in the latter case we have omitted $k=1$ since Figures \ref{fig:size-vs-k} and \ref{fig:resnik-vs-k} clearly demonstrate that $k=1$ seeds perform poorly. It is interesting to note in Figure \ref{fig:resnik-vs-size-no1} that even for the large values of $k$, the Resnik scores {\em increase} with alignment size; we suspect this is because (a) the IID networks are highly similar to each other by construction, and thus (b) our stringent demands on EC1 and EC2 (cf. Table \ref{tab:requirements}) force the networks to align ``properly'', in the sense that the larger the region being aligned, the more ``brittle'' the ``fit'' becomes.

\subsection{Comparison to other methods}
Many other local alignment algorithms exist, but almost none of them seem to work with topology alone. Virtually every paper we've read on local network alignment---and we have read dozens---simply assume that sequence information can and should be used. We have even downloaded and run many of them; the only one we have found that can give anything other than completely null results is AlignMCL \citep{mina2012alignmcl}. Here we summarize their results on the same networks as we used above.

AlignMCL is a local alignment algorithm which uses the popular idea of combining two networks into a single alignment graph and then mining this alignment graph for higher quality local alignments. In order to create this initial alignment graph, AlignMCL utilizes a list of ``orthologs'' created with protein sequence similarity. Then, true to its name, AlignMCL uses a Markov Clustering Algorithm to walk the alignment graph and discover clusters, which it outputs as alignments. In order to adapt AlignMCL to a topology-only setting, we used two different approaches. First, we gave AlignMCL the full list of orthologs between two networks but gave all ortholog pairs a similarity value of 1.0. A second approach, used by \cite{meng2016local}, adapts AlignMCL to use topology only by creating the ``orthologs'' with Orbit Degree Vector similarity (ODV) \citep{Tijana2008} instead of protein sequence similarity.

We ran AlignMCL in several ways against the IID rat-mouse pair. While disallowing sequence similarities was necessary in order to force it to perform topology-only alignment, we found that seeding it only with ODV similarity \citep{Tijana2008}, as did \cite{meng2016local}, performed relatively poorly. We found better results using two methods: first, we adopted the measure used by GRAAL \citep{GRAAL} in which topological similarity used a mix of ODV vector and simple degree difference; we found that using $\alpha=0.8$ for ODVs and $1-\alpha=0.2$ to degree difference gave the best balance. Furthermore, we eliminated degree-1 (leaf) nodes from consideration since both their degree difference of zero, and their high ODV similarity spuriously tricked AlignMCL into aligning leaf nodes to each other essentially at random.

The other confounding factor is that AlignMCL produces many-to-many alignments, while BLANT-extend produces only 1-to-1 alignments. We choose to evaluate it in two ways: either allowing all the non-1-to-1 node pairs, or by forcing 1-to-1 alignments but first {\em sorting by highest Resnik score first}---which is as generous as we could possibly be. The results are depicted in Table \ref{tab:AlignMCL-IID}. As we can see, even when given generous allowances, the Resnik scores are still significantly lower than those depicted in Figures \ref{fig:resnik-vs-k} and \ref{fig:resnik-vs-size-no1}. Run times for AlignMCL on IID rat-mouse varied from a few hours to several days of CPU time, compared to BLANT-extend which we limited to 1 hour of CPU time.

\begin{table}[]
    \centering
    \begin{tabular}{|cc|rc|}
    \hline
    Method  & 1-to-1? &  $n$ & Resnik mean $\pm\sigma$ \\
    \hline
    ODV+Deg & No & 13771 & $4.86\pm 1.98$ \\
    ODV+Deg & Yes &   632 & $7.26 \pm 2.95$ \\
    ", no-deg-1 & No & 10136 & $4.15\pm 2.82$ \\
    ", no-deg-1 & Yes  &   3120     & $6.78\pm 2.96$ \\
    \hline
    \end{tabular}
    \caption{{AlignMCL results applied to IID rat-mouse}: We apply four variations. All four use the best-balance of ODV vector similarity and degree difference (see text); the first two rows allow nodes of degree 1, the last 2 rows do not. Finally, we depict results when we either allow many-to-many mappings (``1-to-1 = No''), or force 1-to-1 mappings by disallowing any node to appear twice, but first {\em sorting Resnik scores highest-to-lowest}; the latter ensures that we give AlignMCL the benefit of the doubt so that each node is given its best possible pairing according to Resnik.}
    \label{tab:AlignMCL-IID}
\end{table}
In our last and most generous test, we seeded AlignMCL with an essentially correct alignment of 15,000 1-to-1 rat-mouse orthologs. After running for 7.3 hours (on the same hardware BLANT-extend used), it correctly returned almost 14,000 out of the 15,000 orthologs. However, they were returned in 7099 distinct and mutually exclusive local alignments, with the largest being just 42 node pairs; there we also 6 alignments with between 20--30 node pairs, 33 having 10--19 node pairs, 3364 with 2-9 pairs, and 3696 ``alignments'' containing only one pair of nodes each.

We believe it is fair to say that using $k=1$ seeds, BLANT-extend compares favourably to AlignMCL, in that our mean and median, and maximum alignment sizes all significantly outperform those of AlignMCL. However, when using $k$-graphlets with $k\ge 8$, BLANT-extend far outperforms AlignMCL, even when AlignMCL is given 1-to-1 orthologs to begin with.

\section{Discussion}
To our knowledge, BLANT (including both the seed and extend elements) is the first and only local network aligner that can produce excellent alignments using topology alone. We have shown that when presented with two large PPI networks of about 15,000 nodes and 300,000 edges each and containing known high topological similarity, BLANT-extend easily finds local alignments of hundreds of node pairs containing near-perfect topological identity. Furthermore, these large local alignments are biologically relevant in that they have very high mean Resnik semantic similarity. We have demonstrated that using $k$-graphlet seeds with 8 or more nodes provides far better opportunity for extending than simply seeding a local alignment with just 1 node pair, even if that node pair is a 1-to-1 ortholog.
We have shown that it soundly outperforms AlignMCL, which \cite{meng2016local} claimed was the best local alignment algorithm they tested.

By preferentially expanding the local alignment outwards using node pairs with high Importance Similarity \citep{HubAlign}, BLANT-extend is able to build large, high quality alignments faster than using an undirected exhaustive search; in all cases we restricted BLANT-extend to 1 hour of run time. (Though not mentioned previously, BLANT-seed takes between 30-60 minutes per network to produce its deterministic $k$-graphlet index, but only seconds to produce the seed pairs once the index exists.)

We believe BLANT can be significantly improved in several ways. BLANT-seed is already written entirely in C; moving BLANT-extend from Python to C will speed its runtime enormously, making it either trivial to run in seconds or minutes, or potentially allowing even larger, high-quality local alignments to be found. 

BLANT seed+extend {\em does} have disadvantages: like BLAST, it requires some amount of {\em identical} topology to exist between two networks before it is able to find seeds worth extending. As a result, it is unable to find good local alignments in networks that are substantially incomplete, noisy, or have little ``common'' edge coverage. Unfortunately this is the case with current BioGRID networks.
While we have verified that results are poor or the full BioGRID networks, we will next attempt similar runs on BioGRID induced on known orthologs. Of course another route to take is to sort node pairs not by Importance, but by sequence similarity. While this defeats the purpose of a topology-only alignment, it may still provide interesting and meaningful results if, for exemple, zero-sequence similarity node pairs were randomly thrown into the sort. This is analogous to what we have recently tried using global network alignment, with excellent results \citep{WeBeat,WangAtkinsonHayesGOpredict}.

\bibliographystyle{natbib}
\bibliography{wayne-all}

\end{document}